\documentclass[12pt]{article}
\usepackage{epsfig,amsfonts,amsthm}
\newcommand{\be}{\begin{equation}}
\newcommand{\ee}{\end{equation}}
\newcommand{\bea}{\begin{eqnarray}}
\newcommand{\eea}{\end{eqnarray}}

\newcommand{\br}{{\bf r }}

\newcommand{\bk}{{\bf k}}

\newcommand{\bp}{{\bf p}}
\newcommand{\bq}{{\bf q}}

\def\lsim{\mathrel{\rlap{\lower4pt\hbox{\hskip1pt$\sim$}}
    \raise1pt\hbox{$<$}}}         

\def\gsim{\mathrel{\rlap{\lower4pt\hbox{\hskip1pt$\sim$}}
    \raise1pt\hbox{$>$}}}         

\title{Non-forward scattering of twisted particles}
\author{Igor~P.~Ivanov
\\
  {\small IFPA, Universit\'{e} de Li\`{e}ge, All\'{e}e du 6 Ao\^{u}t 17, b\^{a}timent B5a, 4000 Li\`{e}ge, Belgium}\\
  {\small Sobolev Institute of Mathematics, Koptyug avenue 4, 630090, Novosibirsk, Russia}\\
  }

\begin{document}

\maketitle

\begin{abstract}
Twisted photons (i.e. photons carrying non-zero orbital angular momentum)
are well-known in optics. Recently, it was suggested to use Compton backscattering
off an ultra-relativistic electron beam to boost optical twisted photons 
into the high energy range. 
However, only the case of strictly forward/backward scattering has been studied so far.
Here, we consider generic kinematic features of processes in which 
a twisted particle scatters with non-zero transverse momentum transfer.
\end{abstract}

\section{Introduction}

In perturbative quantum field theory we assume that interaction among the fields
can be treated as a perturbation of the free field theory.
This perturbation leads to scattering between asymptotically free multiparticle states,
which are usually constructed from the plane wave one-particle states.
This choice greatly simplifies the calculations
and represents a very accurate approximation to the real experimental situation
in virtually all circumstances.
However, one can, in principle, choose any complete basis 
for the one-particle states other than the plane wave basis, provided
that it is still made up of solutions of the free field equations.
Such states can carry new quantum numbers absent in the plane wave choice
and, if experimentally realized, they can offer new opportunities in high-energy physics.

Thanks to the recent experimental progress in optics, it is now possible to create 
laser beams carrying non-zero orbital angular momentum (OAM)
\cite{OAM}, for a recent review see \cite{OAMreview}.
The lightfield in such beams is made of states which are non-plane wave 
solutions of the Maxwell equations. 
Each photon in this lightfield, which we call a {\em twisted photon}, 
carries a non-zero OAM quantized in units of $\hbar$.
Several sets of solutions have been investigated, such as Bessel beams or Gauss-Laguerre beams,
but in all cases the spacial distribution of the lightfield is necessarily 
non-homogeneous in the sense that
the equal phase fronts are not planes but helices.
Such states form a complete basis which can be used to describe the initial and final 
asymptotically free states. Moreover, it is the basis of choice for experimental situations 
when the initial states are prepared in a state of (more or less) definite OAM.

Twisted photons have been produced in various wavelength domains, from radiowave \cite{radio} 
to optical, with prospects to create a brilliant X-ray beam of twisted light in the keV range 
\cite{Xrayproposal}.
Very recently it was suggested to use the Compton backscattering
of twisted optical photons off an ultra-relativistic electron beam to create
a beam of high-energy photons with non-zero OAM \cite{serbo1,serbo2}.
The technology of Compton backscattering is well established, and the 
high-energy electron beams and the OAM optical laser beams are already available.
The suggestion of \cite{serbo1,serbo2} paves the way for the twisted photons, and the twisted states in general, 
into the high-energy physics.
The wealth of new physics opportunities related to this new degree of freedom
is yet to be understood (see \cite{ivanov2} for initial steps in this direction).

In this paper we begin this exploration by focusing on a technical question
of how to calculate the scattering of a one-particle twisted state off plane wave states:
\be
\mbox{twisted} + X(p) \to \mbox{twisted}' + X'(p')\,.\label{general-process}
\ee
Here systems $X$ and $X'$ are one- or multi-particle states described by plane waves,
$p$ and $p'$ being their respective total momenta.
We consider the case of non-forward (and non-backward) scattering, that is, when the
transverse momenta of systems $X$ and $X'$ with respect to the axis used to define the twisted states
are different, $\bp\not = \bp'$.

In this work we focus on the general kinematical properties of the scattering matrix which hold
for all processes of type (\ref{general-process}). 
The purpose of this exercise is twofold. The main goal is to develop the formalism of treating
the scattering processes with twisted particles in the initial and final states.
The secondary goal is to extend the original calculation of \cite{serbo1,serbo2} 
of the strictly backward Compton scattering of twisted optical photons to non-zero transverse momentum transfer. 
The basic questions are: how to write the cross section of this process, 
and how the parameters of the final twisted photon depend on the momentum transfer.

Since the kinematical details turn out to be rather unconventional,
we describe them in a pedagogical manner with the simplest possible process: 
the decay of a massive twisted scalar into two massless scalar.
In this way we separate the effect of non-homogeneous spatial distribution
from the possible effect of non-trivial polarization states, which the twisted photon can have
and which we postpone for a future study.
We stress however that the kinematical features we explore with this example 
are pertinent to all the scattering processes of type (\ref{general-process}).
Additional process-specific properties arise on top of these kinematical features.

The paper is organized as follows.
In Section \ref{section-twisted} we give an introduction into twisted particles and describe some of its properties.
In Section \ref{section-problem} we describe the general problem we tackle: scattering of a twisted particle
accompanied by a non-zero momentum transfer.
We pinpoint the key quantity to analyze, and then, in Section \ref{section-decay}, we investigate this quantity
in detail with a simple pedagogical example.
Section \ref{section-discussion} contain discussion of the results obtained and in the final Section 
we draw our conclusions.
In Appendix we derive some properties of the Bessel functions used in the main text.

\section{Describing twisted states}\label{section-twisted}

\subsection{Spatial distribution}

As mentioned in the introduction, we focus in this paper 
on twisted scalar particles only. In this Section we follow essentially
\cite{serbo1,serbo2}.

We represent a state with non-zero OAM with a Bessel beam-type twisted state.
This is a solution of the wave equation in the cylindric coordinates with definite
energy $\omega$ and longitudinal momentum $k_z$ along a fixed axis $z$, 
definite modulus of the transverse momentum $|\bk|$ (all transverse momenta will be written in bold)
and a definite $z$-projection of OAM.
If the plane wave state $|PW(\bk)\rangle$ is
\be
|PW(\bk)\rangle = e^{-i\omega t + i k_z z} \cdot e^{i\bk\br}\,,
\ee
then a twisted scalar state $|\kappa,m\rangle$ is defined as the following superposition
of plane waves: 
\bea
|\kappa,m\rangle &=& e^{-i\omega t + i k_z z} 
\int {d^2\bk \over(2\pi)^2}a_{\kappa m}(\bk) e^{i\bk \br}\,,\\
\mbox{where}&& a_{\kappa m}(\bk)= (-i)^m e^{im\phi_k}\sqrt{2\pi}{\delta(|\bk|-\kappa)\over \sqrt{\kappa}}\,.
\label{twisted-def}
\eea
In the coordinate space,
\be
|\kappa, m\rangle = e^{-i\omega t + i k_z z} \cdot \psi_{\kappa m}(\br)\,,
\quad \psi_{\kappa m}(\br) = {e^{i m \phi_r} \over\sqrt{2\pi}}\sqrt{\kappa}J_{m}(\kappa r)\,.
\ee
Here, following \cite{serbo1} we call $\kappa$ the conical momentum spread, $m$ is the $z$-projection of OAM,
and the dispersion relation is $k^\mu k_\mu = \omega^2 - k_z^2 - \kappa^2 = M^2$.
We note in passing that the average values of the four-momentum carried by a twisted state is
\be
\langle k^\mu \rangle = (\omega,\, {\bf 0},\, k_z)\,,
\ee
so that $\langle k^\mu \rangle \langle k_\mu \rangle = M^2 + \kappa^2$, which is larger than the true mass of the particle squared.

The transverse spatial distribution is normalized according to
\be
\int d^2\br \psi^*_{\kappa' m'}(\br)\psi_{\kappa m}(\br) = 
\delta_{m,m'} \sqrt{\kappa\kappa'}\int rdr J_{m}(\kappa r) J_{m}(\kappa' r) = \delta_{m,m'}\delta(\kappa-\kappa')\,.
\ee
The plane wave can be recovered from the twisted states as follows:
\bea
|PW(\bk = 0)\rangle & = & \lim_{\kappa \to 0} \sqrt{{2\pi \over\kappa}} |\kappa,0\rangle\,,\label{PWlimit0}\\
|PW(\bk)\rangle & = & \sqrt{{2\pi \over\kappa}} 
\sum_{m=-\infty}^{+\infty} i^m e^{-im\phi_k} |\kappa,m\rangle \,,\quad 
\kappa = |\bk_\perp|\,.\label{PWlimit}
\eea
If needed, these two cases can be written as a single expression:
\be
|PW(\bk)\rangle = \lim_{\kappa \to |\bk|} \sqrt{{2\pi \over\kappa}}
\sum_{m=-\infty}^{+\infty} i^m e^{-im\phi_k} |\kappa,m\rangle\,.
\ee
From these expressions one sees that the twisted states with different $m$ and $\kappa$
represent nothing but another basis for the transverse wave functions.

\subsection{Density of states}

When calculating cross sections and decay rates, we need to integrate the transition
probability over the phase space of the final particles.
When calculating the density of states, we consider large but finite volume 
and count how many mutually orthogonal states with prescribed boundary conditions
can be squeezed inside. In the present case due to the cylindrical symmetry of the problem, 
we choose a cylinder of large radius $R$ and length $L_z$.
In the case of plane waves we have
\be
dn_{PW} = \pi R^2 L_z{dk_z d^2\bk \over (2\pi)^3}\,.
\ee
The full number of states with transverse momenta up to $\kappa_0$ and longitudinal momenta
$|k_z| \le k_{z0}$ is $k_{z0} L_z\cdot R^2\kappa_0^2/4\pi$.

To count the number of twisted states $|\kappa,m\rangle$ in the same volume, 
we specify the boundary condition, e.g. $\psi_{\kappa m}(r=R)=0$, which makes
$\kappa$ discrete such that $\kappa_i R$ is the $i$-th
root of the Bessel function $J_m$. 
Then we note that the position of the first root of the Bessel function $J_m(x)$
is always at $x > m$, and as $m$ grows $x \to m$.
For a given $\kappa$, the maximal $m$ for which the wave can still be contained
inside the cylindrical volume is $m_{max} = \kappa R$,
which has a very natural quasiclassical interpretation. 

If $m$ is small and not growing with $R$, then one can use the well-known
asymptotic form of the Bessel functions to count the number of states:
\be
dn_{tw} = {R d\kappa \, L_z dk_z\, \Delta m\over 2\pi^2}\,.\label{dn_tw1}
\ee
Here, $\Delta m$ is written instead of just 1 to signal the presence of 
a discrete running parameter $m$.

If $m$ is not restricted to small values, this asymptotic form 
of $J_{m}(x)$ cannot be used since it requires $m^2 \lsim x$. 
Instead, the so-called approximation by tangents can be used, 
which gives the following density of states:
\be
dn_{tw} = \sqrt{m_{max}^2-m^2} {d\kappa \over \kappa}{\Delta m \over \pi}{L_z dk_z \over 2\pi}\,.
\ee
In the limit $m \ll m_{max}\equiv \kappa R$ this expression reproduced (\ref{dn_tw1}).
Alternatively, one can calculate the radial part of the density of states
via the adiabatic invariant as suggested in \cite{serbo2}.
The number of radial excitations $n_r$ for a fixed $m$ is
\be
n_r = \int_{m/\kappa}^{R}{k_r(r)dr \over \pi}\,,\quad k_r(r) = \sqrt{\kappa^2 - {m^2\over r^2}}\,. 
\ee
The density of states is then given by
\be
dn_r = {dn_r \over d\kappa} d\kappa = \sqrt{m^2_{max}-m^2}{d\kappa \over \kappa \pi}\,.
\ee

One important remark is in order.
Effectively, switching from the plane wave to twisted state basis for the final particles
implies replacement
\be
d^2\bk \to 4\sqrt{1-{m^2 \over m_{max}^2}}\, \kappa d\kappa\, {\Delta m \over m_{max}}\,.
\ee
Note that the contribution of each ``partial wave'' with a fixed $m$ vanishes in the infinite volume limit as $1/R$.
However, the number of partial waves grows $\propto R$, and in order to get a non-vanishing
result for a physical observable, one must integrate over the full available $m$ interval up to $m_{max}$.
This holds even if the transverse momenta stay small, and is related to the fact that 
the plane wave contains contributions from all impact parameters with respect to any axis non-collinear
to its momentum.

Another expression one needs for the probability calculations is the
normalization constants for the one-particle states. 
A usual plane wave one-particle state is normalized to $2E\cdot V$;
to renormalize it to one particle per the entire volume, the plane wave
should be multiplied by $N_{PW}$, with 
\be
N^2_{PW} = {1 \over 2E V}\,,\quad V = \pi R^2 L_z\,.
\ee
For a twisted states the corresponding normalization factor $N_{tw}$ is
\be
N_{tw}^2 = {1 \over 2E}{\pi \kappa\over \sqrt{m_{max}^2-m^2} L_z}\,,
\ee
which in the small-$m$ case simplifies to 
\be
N_{tw}^2 \approx {1 \over 2E} {\pi \over R L_z}\,,
\ee
which was also derived in \cite{serbo2}.
Note however that even in the general case
the product of the normalization constant squared and the density of states for each final
twisted particle is simplified as
\be
N_{tw}^2 dn_{tw} = {d\kappa dk_z \Delta m \over 2E \cdot 2\pi}\,.
\ee

\section{Non-forward scattering of a twisted state: generic features}\label{section-problem}

\begin{figure}[!htb]
   \centering
\includegraphics[width=8cm]{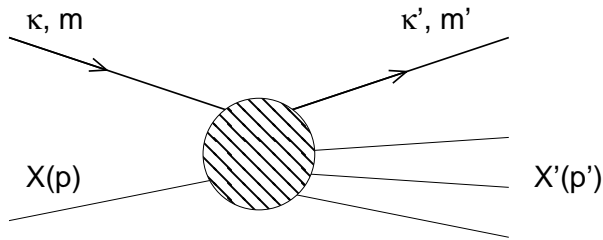}
\caption{A generic process of non-forward scattering of a twisted state off plane waves.}
   \label{fig1}
\end{figure}

Consider again the generic process (\ref{general-process}) which is schematically depicted in Fig.~\ref{fig1}.
Here we have one twisted particle in the initial and one in the final states.
Since we deal here with the kinematical features of the process,
it is inessential whether these particles are of the same type or not.
These two particles are described by the states $|\kappa,m\rangle$ 
and $|\kappa',m'\rangle$ defined with respect to a common $z$-axis. 
All the other particles are assumed to be plane waves, which allows us to define
the total transverse momenta $\bp$ and $\bp'$ of the systems $X$ and $X'$, respectively. 
The transverse momentum transfer, $\bq\equiv \bp-\bp'$ is also well-defined, and
we can consider two cases: (i) strictly forward scattering, $\bq = 0$,
(2) non-forward scattering, $\bq \not = 0$. Note that we use the term ``forward'' 
for any process with zero transverse momentum transfer, i.e. both for strictly forward 
and strictly backward scattering.

Suppose that we know the $S$-matrix for the same process with {\em all} particles 
represented by the plane waves. If the would-be twisted particles in the initial and final states 
have momenta $k$ and $k'$, the plane wave $S$-matrix has the familiar form:
\be
S_{PW}=i (2\pi)^4\delta^{(4)}(k+p- k' - p')\cdot {\cal M}(\bk^2,\bk^{\prime 2},\phi_{kk'}).\label{S_PW}
\ee
The invariant amplitude ${\cal M}$ depends among other on the transverse momenta
squared $\bk^2$ and $\bk^{\prime 2}$ as well as on the azimuthal angle $\phi_{kk'}$ between them,
which is explicitly indicated in (\ref{S_PW}). 
Any other azimuthal angle, say, between $\bk$ and $\bp$, can be written
as $\phi_{kp} = \phi_{kq}+\phi_{qp}$, and therefore it is expressible via $\phi_{kk'}$
and internal angles in the systems $X$ and $X'$. 

To get the $S$-matrix for the twisted particle scattering process (\ref{general-process}),
one just needs to apply (\ref{twisted-def}) to the initial and final states \cite{serbo1}:
\be
S_{tw}=\int {d^2\bk \over(2\pi)^2} {d^2\bk' \over(2\pi)^2}
a^*_{\kappa' m'}(\bk') a_{\kappa m}(\bk)\, S_{PW} \,.\label{S_tw} 
\ee
The delta-functions present in (\ref{S_tw}) fix $\bk^2 = \kappa^2$,
$\bk^{\prime 2} = \kappa^{\prime 2}$, and, as we will see later,
they specify the angle $\phi_{kk'}$ up to its sign.
If the plane wave matrix element is not sensitive to the sign of the angle $\phi_{kk'}$, 
${\cal M}(\kappa^2,\kappa^{\prime 2},\phi_{kk'})= {\cal M}(\kappa^2,\kappa^{\prime 2},-\phi_{kk'})$, 
then it can be taken out of the integral,
and we are left with the following transverse process-independent master integral:
\be
{\cal I}_{m,m'}(\kappa,\kappa',\bq) = \int {d^2\bk \over(2\pi)^2} {d^2\bk' \over(2\pi)^2}
a^*_{\kappa' m'}(\bk') a_{\kappa m}(\bk) \delta^{(2)}(\bk - \bk' +\bq)\,.\label{master}
\ee
If the matrix element depends on the sign of $\phi_{kk'}$, then it can be decomposed
into a symmetric and antisymmetric terms
\be
{\cal M}_s = {{\cal M}(\phi_{kk'})+ {\cal M}(-\phi_{kk'}) \over 2}\,,\quad
{\cal M}_a = {{\cal M}(\phi_{kk'})- {\cal M}(-\phi_{kk'}) \over 2}\,,
\ee
then in the twisted scattering amplitude ${\cal M}_s$ will be accompanied by the master integral (\ref{master}),
while ${\cal M}_a$ will enter together with a slightly modified version of the 
master integral, $\bar {\cal I}_{m,m'}(\kappa,\kappa',\bq)$, 
in which the integrand of (\ref{master}) is multiplied by the sign function
$\epsilon(\phi_{kk'})$. In fact, two interfering plane wave amplitudes appearing here
might lead to novel observable effects in twisted particle cross sections,
see details in \cite{ivanov2}.

In the next Section we compute the master integral (\ref{master}) 
and explore the singularities of $|{\cal I}|^2$
which appear in the cross section/decay rate calculations.
However, let us briefly summarize the results right away:
\begin{itemize}
\item
In the strictly forward case, $\bq=0$, ${\cal I}_{m,m'} \propto \delta(\kappa-\kappa')\delta_{m,m'}$,
that is, the quantum numbers of the twisted state are transferred to the final particles
without any change.
\item
In the non-forward case, $\bq \not = 0$, a distribution over $\kappa'$ arises, 
and $\kappa'$ can be large if the momentum transfer is large.
\item
At any non-zero momentum transfer instead of $\delta_{m,m'}$ we observe
a distribution of $m'$ over the entire possible range of values, $-m'_{max} \le m' \le m'_{max}$,
where $m'_{max} = \kappa'R \to \infty$.
\item
$|{\cal I}|^2$ contains singularities at the border of the kinematically allowed region,
which must be carefully dealt with.
\end{itemize}

\section{Decay of a twisted scalar}\label{section-decay}

We would like to explore the integral (\ref{master}) in the context of the simplest possible problem:
decay of a twisted scalar particle with mass $M$ into a pair of massless distinguishable particles
due to the cubic interaction $g\cdot \Phi\phi_1\phi_2$.
To make the presentation more pedagogical, we will first calculate the decay rate 
when both particles in the final state are plane waves, 
then for the plane wave plus twisted final state, and finally
for the case when both final particles are twisted.
We will calculate the decay width in the center of mass frame defined by $k_z=0$. 
This is not the true rest frame because due to the transverse motion 
a twisted particle is never at rest. 

\subsection{Two plane waves}

For future reference, let us first recall the calculation for the standard case
when all the particles including the initial one are plane waves.
The $S$-matrix is given as usual by
$S = i(2\pi)^4\delta^{(4)}(p-k_1-k_2)\cdot g$.
When squaring the delta-function, we use the standard prescription
\be
[\delta^{(4)}(p-k_1-k_2)]^2 \longrightarrow \delta^{(4)}(p-k_1-k_2)\cdot \delta^{(4)}(0)
= \delta^{(4)}(p-k_1-k_2) {VT \over (2\pi)^4}\,.
\ee
We also use the plane wave normalization for all the particles,
The decay probability per unit time for a particle at rest is:
\bea
d\Gamma &=& {(2\pi)^4 g^2 \delta^{(4)}(p-k_1-k_2) VT \over T}\cdot (N_{PW}^2)^3
\cdot dn_{PW}(k_1) dn_{PW}(k_2) \nonumber\\
&=& {g^2\over (2\pi)^2} {\delta(M - \omega_1-\omega_2) \over 8M\omega_1\omega_2} d^3k_1\,,\label{PWdecaydiff}
\eea
so that the total width is
\be
\Gamma = {g^2 \over 16\pi M}\,.\label{PWdecay}
\ee

Now we repeat this calculation for the initial twisted state $|\kappa,m\rangle$, 
while keeping the plane wave basis the final particles.
The $S$-matrix is 
\bea
S &=& i(2\pi)^4 g\,\delta(E-\omega_1-\omega_2)\delta(k_{1z}+k_{2z}) 
\int {d^2 \bk \over (2\pi)^2} a_{\kappa m}(\bk) \delta^{(2)}(\bk-\bk_1-\bk_2)
\label{Stwisted1}\\
&=& i(2\pi)^4 g\, \delta(E-\omega_1-\omega_2)\delta(k_{1z}+k_{2z})
{(-i)^m \over(2\pi)^{3/2}} e^{im\phi_{12}}{ \delta(\kappa-k_{12}) \over \sqrt{\kappa}} \,,
\eea
where $k_{12} \equiv \sqrt{\bk_1^2 + \bk_2^2 + 2|\bk_1| |\bk_2|\cos(\phi_1-\phi_2)}$
and $\phi_{12}$ is the angle of the 2D vector $\bk_{12}$ w.r.t. some axis $x$. 
The phase factor is inessential and $m$ disappears in the decay rate.

When squaring the above expression, we encounter the square of $\delta(\kappa-k_{12})$, which is treated
as in \cite{serbo1,serbo2}:
\be
\left[\delta(\kappa-k_{12})\right]^2 = \delta(\kappa-k_{12})\delta(0) \to \delta(\kappa-k_{12}){R\over \pi}\,.
\label{delta-trick}
\ee
This prescription comes from the observation that at large but finite $R$ and at $\kappa=k_{12}$ 
the divergent integral in (\ref{Stwisted1}) is regularized as
\be
\delta(0) = \int_0^\infty rdr [J_{m}(\kappa r)]^2 \to 
\int_0^R rdr [J_{m}(\kappa r)]^2 \approx {R \over \pi}\,.
\ee
Thus, with the normalization factors plugged in, the decay rate has the form
\bea
d\Gamma &=& {(2\pi)^3 g^2 \over 8 E \omega_1\omega_2 \cdot T}\delta(E-\omega_1-\omega_2)\delta(k_{1z}+k_{2z})T L_z 
{\delta(\kappa-k_{12})\over \kappa}{R\over \pi} \cdot {1\over V^2} {\pi \over RL_z}\cdot 
{Vd^3k_1 \over (2\pi)^3} {Vd^3k_2 \over (2\pi)^3}\nonumber\\
&=& {g^2 \over (2\pi)^3 } {\delta(\kappa-k_{12}) \over \kappa}{\delta(E-\omega_1-\omega_2) \over 8 E \omega_1\omega_2 }\, dk_{z}
\,d^2\bk_1\, d^2\bk_2\,.
\eea
For the transverse integral we write
\be
\int d\phi_2 \,{\delta(\kappa-k_{12}) \over \kappa} = 
2 \int d\phi_2 \,\delta\left[\kappa^2 - \bk_1^2-\bk_2^2-2|\bk_1||\bk_2|\cos(\phi_1-\phi_2)\right] = {1 \over \Delta}\,,
\label{angular-integral}
\ee
where
\be
\Delta = {1\over 4}\sqrt{2(\bk_1^2\bk_2^2+\kappa^2 \bk_1^2+\kappa^2 \bk_2^2)
- \kappa^4 - \bk_1^4 - \bk_2^4}\label{Delta-def}
\ee
is the area of the triangle with sides $\kappa$, $|\bk_1|$ and $|\bk_2|$.
\begin{figure}[!htb]
   \centering
\includegraphics[width=6cm]{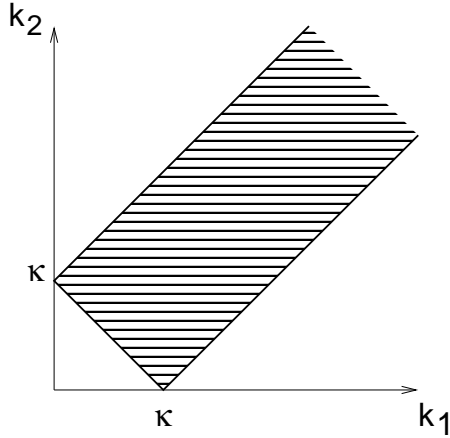}
\caption{The allowed kinematical region of the values of $|\bk_1|$ and $|\bk_2|$ 
for a fixed $\kappa$ defined by the ``triangle rules'' (\ref{triangle-rule}).}
   \label{fig-region}
\end{figure}
The angular integral can be non-zero only if such a triangle can be formed, that is
if $|\bk_1|$ and $|\bk_2|$ satisfy the ``triangle rules'':
\be
\kappa \le |\bk_1|+|\bk_2|\,,\quad
|\bk_1| \le \kappa+|\bk_2|\,,\quad
|\bk_2| \le \kappa+|\bk_1|\,.\label{triangle-rule}
\ee
The allowed values of $|\bk_1|$ and $|\bk_2|$ for a fixed $\kappa$ are shown in Fig.~\ref{fig-region}.
Note that the angular integral (\ref{angular-integral}) receives equal contributions from two
points:
\be
\phi_1-\phi_2 = \pm \delta_{12}\,,\quad \mbox{where} \quad
\delta_{12} = \arccos \left({\kappa^2 - \bk_1^2-\bk_2^2 \over 2|\bk_1||\bk_2|}\right)\,.
\ee
As usual, the energy delta function can be killed by the $k_{z}$ integration
\bea
\int dk_{z} {\delta(E-\omega_1-\omega_2) \over \omega_1\omega_2} = {2 \over E k_z^*}\,,
\eea
where 
\be
k_z^* = {1 \over 2E}\sqrt{E^4+\bk_1^4+\bk_2^4 - 2(E^2\bk_1^2 + E^2 \bk_2^2 + \bk_1^2 \bk_2^2)}\,.
\ee
The decay rate becomes
\be
d\Gamma = {g^2 \over 4\pi^2} {1 \over 4 E^2 k_z^*} {|\bk_1| d|\bk_1|\, |\bk_2| d|\bk_2|\over \Delta} \,.\label{dGamma-PWPW}
\ee
The integration region over $|\bk_1|$ and $|\bk_2|$ is defined by the requirement that, in addition to (\ref{triangle-rule}), 
the longitudinal momentum $k_z^*$ is well-defined, which cuts the rectangular shape shown in Fig.~\ref{fig-region} from above.
It is conveniently described with variables 
\be
x = {|\bk_1|-|\bk_2| \over \kappa}\,,\quad z = {|\bk_1|+|\bk_2| \over \kappa}\,,
\quad x\in [-1,1]\,,\quad z\in [1,z_{max}]\,,\quad z_{max} \equiv {E\over \kappa} > 1\,.\label{limits-xz}
\ee 
In these variables 
\be
\Delta = {\kappa^2 \over 4}\sqrt{(z^2-1)(1-x^2)}\,,\quad
k_z^* = {E\over 2}\sqrt{\left(1-{x^2 \over z_{max}^2}\right)\left(1-{z^2 \over z_{max}^2}\right)}\,,
\ee
and the decay rate takes form
\be
d\Gamma = {g^2 \over 16 \pi^2 E}\cdot {(z^2-x^2)dzdx \over \left[(z_{max}^2-z^2)(z_{max}^2-x^2)(z^2-1)(1-x^2)\right]^{1/2}}\,.
\ee
This integral can be taken exactly, 
and it gives the decay width of the twisted scalar particle
\be
\Gamma = {g^2 \over 16\pi E}\,,\label{twisted-decay}
\ee
which is a very natural result. In the limit $\kappa \to 0$, we recover the plane wave 
decay width (\ref{PWdecay}).

\subsection{Twisted state plus plane wave}

Let us now describe the final state as twisted state plus a plane wave:
\be
|\kappa,m\rangle \to |\kappa_1,m_1\rangle + |PW(\bk_2)\rangle\,.
\ee
Since the full decay width cannot depend on the basis we choose for the final particles,
we must recover the same result (\ref{twisted-decay}) in this basis.
In addition to that, we also want to know how the final twisted state parameters $\kappa_1$ and $m_1$
are related to the initial state parameters $\kappa$ and $m$. 

\subsubsection{Forward case}

The $S$-matrix is now expressed in terms of the master integral (\ref{master}):
\be
S = i (2\pi)^4 g \delta(E-\omega_1-\omega_2)\delta(k_{z1}+k_{z2}) 
\cdot {\cal I}_{m,m_1}(\kappa,\kappa_1,\bk_2)\,.
\ee
Let us first calculate this integral in the strictly forward case $\bk_2=0$:
\bea
{\cal I}_{m,m_1}(\kappa,\kappa_1,0) &=& 
{i^{m_1-m} \over(2\pi)^3} \int d^2\bk\, d^2\bk_1 e^{im\phi-im_1\phi_1}{\delta(|\bk|-\kappa) \over\sqrt{\kappa}}
{\delta(|\bk_1|-\kappa_1) \over\sqrt{\kappa_1}} \delta^{(2)}(\bk-\bk_1)\nonumber\\
&=& {i^{m_1-m} \over(2\pi)^3}\sqrt{\kappa\kappa_1}\int d\phi\, d\phi_1\, e^{im\phi - im_1\phi_1}\, \cdot
2\delta(\bk^2-\bk_1^2)\delta(\phi-\phi_1)\nonumber\\
&=& {1 \over(2\pi)^2}\delta(\kappa-\kappa_1) \delta_{m,m_1}\,,\label{master-forward}
\eea
which was first obtained in \cite{serbo1}.
This result implies that the twisted state quantum number are transferred from the initial
to the final twisted particle without any change.
The differential decay rate is
\bea
d\Gamma &=& {g^2 \over (2\pi)^2}\, {\delta(E-\omega_1-\omega_2)\delta(k_{1z}+k_{2z}) \over 8E\omega_1\omega_2}
\delta_{m,m_1}\delta(\kappa-\kappa_1)\cdot d\kappa_1 dk_{1z} d^3k_2\nonumber\\
&=& {g^2 \over (2\pi)^2} {\delta(E-\omega_1-\omega_2) \over 8 E \omega_1 \omega_2 } d^3 k_2
= {g^2 \over (2\pi)^2} {d^2\bk_2 \over 4 E^2 k_z^*}\,,\label{result-strictly-forward}
\eea
and we stress that this result is applicable only at $\bk_2 =0$.

\subsubsection{Non-forward case: the master integral}

Now we consider the non-forward case. The master integral can be rewritten as
\be
{\cal I}_{m,m_1}(\kappa,\kappa_1,\bk_2) = {i^{m_1-m} \over(2\pi)^3}\sqrt{\kappa\kappa_1}\int d\phi d\phi_1\, 
e^{im\phi - im_1\phi_1}\, \cdot \delta^{(2)}(\bk-\bk_1-\bk_2)\,.\label{master2}
\ee
There are two ways to look at this integral. 
First, we can rewrite 
\be
\delta^{(2)}(\bk-\bk_1-\bk_2) = {1\over(2\pi)^2} \int d^2\br\,  
e^{i\br\bk - i \br \bk_1 - i \br\bk_2}
\label{method1start}
\ee
and represent the integral as 
\be
{\cal I}_{m,m_1}(\kappa,\kappa_1,\bk_2)  = {i^{m_1-m} e^{i(m-m_1)\phi_2}\over(2\pi)^2}\sqrt{\kappa\kappa_1}
\cdot 
\int_0^\infty rdr J_m(\kappa r) J_{m_1}(\kappa_1 r) J_{m-m_1}(|\bk_2| r)\,,\label{master3}
\ee
where $\phi_2$ is the azimuthal angle of $\bk_2$.
Note that although we are considering the case with two twisted particles (one in the initial and one in the final state),
an integral over three Bessel function arises automatically.

The master integral can be also calculated by a direct integration over angles in (\ref{master2}).
We rewrite the delta-function as
\bea
\delta^{2}(\bk-\bk_1- \bk_2 ) &=& 2 \delta[\kappa^2 - (\bk_1 + \bk_2)^2]
\,\delta(\phi - \phi_{\bk_1+\bk_2}) \nonumber\\
&=& 2 \delta[\kappa^2 - \kappa_1^2 - \bk_2^2 - 2 \kappa_1 |\bk_2| \cos(\phi_1-\phi_2)]
\,\delta(\phi - \phi_{\bk_1+\bk_2}) \,.
\eea
One sees that the integral can be non-zero only if it is possible
to form a triangle with sides $\kappa$, $\kappa_1$, and $|\bk_2|$;
thus, $\kappa_1$ and $|\bk_2|$ satisfy the same triangle rules as in (\ref{triangle-rule}). 
Fixing $\kappa_1$ and $|\bk_2|$ means that the triangle can be formed only at
two choices of relative azimuthal angles of $\bk$, $\bk_1$ and $\bk_2$.
Let us introduce
\be
\delta_1 = \arccos\left({\kappa^2+\kappa_1^2 - \bk_2^2 \over 2 \kappa\kappa_1}\right)\,,
\quad
\delta_2 = \arccos\left({\kappa^2+\bk_2^2 - \kappa_1^2 \over 2 \kappa |\bk_2|}\right)\,,\label{deltas}
\ee
Then, the two configurations of transverse momenta correspond to
\be
\phi -\phi_1 = \pm \delta_1\,,\quad \phi -\phi_2 = \mp \delta_2\,,
\ee
so that the signs of $\phi -\phi_1$ and $\phi -\phi_2$ are always opposite.
The master integral then becomes
\be
{\cal I}_{m,m_1}(\kappa,\kappa_1,\bk_2) = {i^{m_1-m} \over(2\pi)^3}\sqrt{\kappa\kappa_1}
e^{i(m-m_1)\phi_2}{\cos[m_1\delta_1 - (m-m_1)\delta_2] \over \Delta}\,,\label{master4}
\ee
where $\Delta$ is the same as in (\ref{Delta-def}) with $\bk_1^2$ replaced by $\kappa_1^2$.
Comparison of (\ref{master3}) with (\ref{master4}) gives the result for the integral
of the triple Bessel function product. See also \cite{JJJ} for some mathematics
involved in evaluation of this and similar integrals.

Let us also check the $\bk_2\to 0$ limit of the result (\ref{master4}).
When $|\bk_2|\ll \kappa$, the distribution over $\kappa_1$ spans from
$\kappa - |\bk_2|$ to $\kappa + |\bk_2|$, see Fig.~\ref{fig-region}. The angle $\delta_1 \to 0$,
while $\delta_2$ can still be arbitrary. However, in the $\bk_2\to 0$ limit
only $m-m_1=0$ term survives due to $J_{m-m_1}(|\bk_2| r)$, 
so that the cosine in (\ref{master4}) approaches unity.
The analysis of $\Delta$ shows that 
\bea
\lim_{|\bk_2|\to 0} { \sqrt{\kappa\kappa_1} \over \Delta} = 
2\pi \delta(\kappa-\kappa_1)\,,\label{limitk2}
\eea
so that one indeed recovers the strictly forward result for the master integral
given in (\ref{master-forward}). Note that although we wrote simply $|\bk_2|\to 0$,
the exact definition of this limit is involved and is discussed in Section~\ref{section-m1}.

Note also that the modified master integral $\bar {\cal I}_{m,m'}(\kappa,\kappa',\bq)$
mentioned in the previous Section has the same form as (\ref{master4}) but
contains sine instead of cosine of $m_1\delta_1 + (m-m_1)\delta_2$. 
This modified integral does not enter our particular calculation since 
the amplitude is symmetric under the $\phi_{kk'}$ sign flip.

\subsubsection{Non-forward case: squaring the amplitude}

After integrating over $k_z$, the decay rate can be written as
\be
d\Gamma = 4\pi^3 g^2 {1 \over 4 E^2 k_z^*} \cdot 
|{\cal I}_{m,m_1}(\kappa,\kappa_1,\bk_2)|^2 \cdot {d\kappa_1 \over R}\, d^2\bk_2\,.\label{dGamma_twPW0}
\ee
Note that this decay rate is differential not only in $\kappa_1$ and $\bk_2$ but also in the discrete variable $m_1$;
the full decay width includes integrals over momenta and a summation over all possible $m_1$'s.

A close inspection shows that the immediate integration over $\kappa_1$ or $\bk_2$ 
cannot be done due to singularities of $|{\cal I}|^2$ along the boundaries of the kinematically allowed
region shown in Fig.~\ref{fig-region}.
In contrast to the plane wave case, the denominator now contains $\Delta^2$ instead of just $\Delta$.
Therefore, in terms of variables $x$ and $z$ one encounters singularities of the form
$$
\int_{-1}^1 {dx \over 1-x^2} \quad \mbox{and} \quad \int_1^{z_{max}} {dz \over z^2-1}\,.
$$
Clearly, this is an artefact of the infinite radial integration range. 
If instead we take $R$ to be large but finite,
we expect that a trick similar to (\ref{delta-trick}) should be at work, 
namely that after regularization $|{\cal I}|^2$ 
would yield $R$ times a less singular function.

This trick does not seem to work for each $m_1$ separately.
However, as we prove in Appendix, it works for $|{\cal I}|^2$ summed over all
possible $m_1$. In the limit $R\to \infty$ we obtain
\be
\sum_{m_1=-\infty}^{+\infty} |{\cal I}_{m,m_1}(\kappa,\kappa_1,\bk_2)|^2
= {1 \over (2\pi)^5} {R\kappa_1 \over \pi} {1 \over \Delta}\,,
\ee
with the same $\Delta$ as before. The regularization parameter $R$ then disappears
from the result, and the decay rate reads
\be
d\Gamma = {g^2 \over 4\pi^2} {1 \over 4 E^2 k_z^*} \cdot 
{\kappa_1 d\kappa_1\, |\bk_2| d |\bk_2|\over \Delta}\,.\label{dGamma_twPW}
\ee
Comparing (\ref{dGamma_twPW}) with the previous results (\ref{result-strictly-forward}) 
and (\ref{dGamma-PWPW}) leads us to two conclusions.
\begin{itemize}
\item
The transition from the strictly forward to the non-forward cross section/decay rate
consists in replacement
\be
{\delta(\kappa-\kappa_1) \over \kappa_1} \to {1 \over 2\pi\Delta}\,.\label{the_rule}
\ee
\item
The result (\ref{dGamma_twPW}) for a twisted particle in the final state
coincides with the result (\ref{dGamma-PWPW}) for the case when both final particles
are plane waves. 
\end{itemize}
Although these conclusions were drawn for the specific process we consider,
the way it is derived suggests that this might be a universal feature
for many (or all) processes of type (\ref{general-process}).

\subsection{Two twisted states}

For completeness, let us also recalculate the decay rate in the basis when
both final particles are described by twisted states: $|\kappa_1,m_1\rangle$ and $|\kappa_2,m_2\rangle$.
We remind that all twisted states are defined with respect to the same common $z$ axis.
It turns out that this calculation closely follows the case of 
twisted state plus plane wave just considered. This is not surprising because
the appearance of the triple Bessel integral highlights the fact that when two particles
(one in the initial and one in the final state) are twisted, the third one is automatically
projected from the plane wave onto an appropriately defined twisted state as well.

The $S$-matrix takes the form 
\be
S = i g (2\pi)^{3/2}\delta(E-\omega_1-\omega_2)\delta(k_{z1}+k_{z2}) \cdot \delta_{m,m_1+m_2}
\sqrt{\kappa\kappa_1\kappa_2} \int rdr\, J_m(\kappa r)J_{m_1}(\kappa_1 r) J_{m_2}(\kappa_2 r)\,,
\ee
and we again encounter the triple Bessel-function integral. 
The decay rate is written as
\be
d\Gamma = {g^2 \over 4\pi^2} {1 \over 4 E^2 k_z^*} \cdot 
{\kappa_1 d\kappa_1\, \kappa_2 d \kappa_2\over \Delta}\,.\label{dGamma_twtw}
\ee
This expression is identical to (\ref{dGamma_twPW}) up to the obvious replacement $|\bk_2| \to \kappa_2$;
its integration over all conical momenta spreads $\kappa_1$, $\kappa_2$ gives again (\ref{twisted-decay}).

\section{Discussion}\label{section-discussion}

\subsection{Distribution over $\kappa_1$}

Let us first discuss what typical values of $\kappa_1$ and $m_1$ essentially contribute
to the decay rate. The differential decay rate (\ref{dGamma_twPW}) 
shows that at large $|\bk_2|\gg \kappa$ the conical momentum spread 
$\kappa_1$ is limited to the interval from $|\bk_2|-\kappa$ to $|\bk_2|+\kappa$ with an inverse square root
singularity at the endpoints. This singularity is integrable, so that the entire interval more or less equally 
contributes to the integral. Since $|\bk_2|$ can be as high as $E$, 
the total decay width is therefore dominated by large $\kappa_1 \gg \kappa$.
In a more complicated process, the decay rate or the cross section
will include the amplitude squared which can serve as a cut-off function.
For example, in the Compton scattering one expects that $\kappa_1$ up to $\sim m_e$ will
contribute to the cross section.

\subsection{Distribution over $m_1$}\label{section-m1}

Although we found a result for the decay rate summed over all $m_1$, 
we can trace the main $m_1$-region from the intermediate formulas. The result
is that essentially all $m_1$ from minus to plus infinity are important for the decay rate,
which is in a strong contrast to the strictly forward result $m=m_1$.

Indeed, the distribution over final $m_1$ can be seen in (\ref{dGamma_twPW0}), where
the master integral ${\cal I}_{m,m'}$ is given by (\ref{master4}). 
For generic transverse momenta, growth of $m_1$ leads to oscillations of the cosine function
with constant amplitude. Thus, when averaging over the entire $\kappa_1$ interval,
one can approximate cosine squared by $1/2$, and the dependence on $m_1$ drops off.
This result holds for any non-zero transverse momentum transfer $|\bk_2|\to 0$.

Since the forward and non-forward $m_1$-distributions are so dramatically different,
a natural question arises whether there is a continuous transition from 
the non-forward to the forward scattering. The answer to this question involves an accurate treatment
of two limits: $|\bk_2|\to 0$ and $R \to \infty$. Let us keep $R$ large but finite, and set $|\bk_2|\to 0$;
then the transition is smooth.
Looking at the triple-Bessel representation of the master integral (\ref{master3}) with the upper limit
replaced by $R$, one sees that the result will begin to significantly decrease only
when the position of the first node of the last Bessel function $J_{m-m_1}(|\bk_2|r)$
falls outside of the integration range, that is for $|\bk_2| \lsim |m-m_1|/R$, where $m_1 \not = m$.
If $m_1=m$, then at $|\bk_2| \ll 1/R$ the last Bessel function can be approximated by the unity.
Therefore, the limit $|\bk_2|\to 0$ taken in (\ref{limitk2}) implies that
$$
|\bk_2|\to 0\quad \mbox{and}\quad R \to \infty\quad \mbox{provided that}\quad  |\bk_2|R \ll 1\,.
$$
If instead $R\to \infty$ at fixed $|\bk_2|$, then the transition of non-forward to forward
results is discontinuous at $|\bk_2|=0$.

In \cite{serbo1} it is claimed with the specific example of the Compton
cross-section that if the transverse momentum transfer is small compared to $\kappa$ 
(in our notation, finite $|\bk_2| \ll \kappa$ at infinite $R$),
then the $m_1$-dependence has a narrow distribution peaked at $m_1=m$.
Our analysis shows that this cannot be true, and this conclusion of \cite{serbo1}
is likely to be the result of an incorrect small-$|\bk_2|$ approximation for 
${\cal I}_{m,m_1}(\kappa,\kappa_1,\bk_2)$.

\subsection{Orbital angular momentum vs. orbital helicity}

Looking back at the formalism used, we can conclude that 
our result that all $m_1$ essentially contribute to the decay rate/cross section
just reflects the unfortunate choice of the same common axis $z$ for
all the twisted states appearing in the process. 
It does not give a clue of how twisted the final particles are with respect to their own 
propagation axes defined by their average values of the 3-momentum operator. 
Indeed, even a simple non-forward plane wave when expanded in the basis of twisted states
contains all partial waves, see (\ref{PWlimit}). Nevertheless it carries a zero
orbital angular momentum with respect to its own direction of propagation.
Therefore, it appears that a more physically reasonable
quantity is the {\bf ``orbital helicity''}, projection of orbital angular momentum 
on the axis of motion. The relevant question is then how this
``orbital helicity'', not the OAM with respect to a fixed axis, is transferred
from the initial to the final twisted state.
We postpone this question for future studies.

\section{Conclusions}\label{section-conclusions}

Photons carrying non-zero orbital angular momentum (twisted photons) are well known in optics.
Thanks to the recent suggestion \cite{serbo1,serbo2} to use the Compton backscattering of optical twisted photons 
off a high-energy electron beam, twisted photons are now entering the high-energy physics.
If the new degree of freedom they offer is effectively realized in experiment,
twisted photons can bring novel opportunities to particle physics.

In this paper we took a first look at the non-forward scattering of a twisted particle 
off a system of plane waves:
$$
\mbox{twisted} + X(p) \to \mbox{twisted}' + X'(p')\,,\quad \bp' \not = \bp\,,
$$
and investigated universal kinematical features pertinent to such processes.

With the simple example of the decay of a scalar twisted particle, 
we evaluated the master integral appearing in such processes, discussed
its singularities and explained how to write the cross section/decay rate.
These results can be now used, for example, to investigate the Compton
backscattering of the twisted photons in the non-forward region, which was missing in
the original suggestion \cite{serbo1,serbo2}.

Discussing the results obtained, we came to the conclusion that
a more physically motivated quantity to describe a twisted state
would be the orbital angular momentum projection not on the 
reaction axis but on the direction of the outgoing twisted particle. 
Incorporation of this ``orbital helicity'' into the present formalism 
remains to be done.

\section*{Acknowledgements} 
The author is grateful to V.~Serbo for numerous useful discussions on this subject
and to I.~Ginzburg for comments.
This work was supported by the Belgian Fund F.R.S.-FNRS via the
contract of Charg\'e de recherche, and in part by grants
RFBR No.08-02-00334-a and NSh-3810.2010.2.

\appendix

\section{Regularization of $|{\cal I}|^2$}

Here we calculate the large-$R$ behavior of the 
$m_1$-sum of the squares of the triple-Bessel integral:
\be
\sum_{m_1=-m_{max}}^{+m_{max}} \left[\int_0^R rdr J_m(\kappa r)J_{m_1}(\kappa_1 r) J_{m-m_1}(\kappa_2 r)\right]^2\,,
\label{integral-appendix}
\ee
which appears in the decay rate (\ref{dGamma_twPW0}).
Evaluation of the integral itself with $R\to \infty$ performed in the main text
shows that it can be non-zero only if $\kappa$, $\kappa_1$, $\kappa_2$ satisfy the triangle rules (\ref{triangle-rule}),
i.e. a triangle with these sides can be constructed.
Since $m$ describes the initial state, we take it small and not growing with $R$: 
$m\ll m_{1max} = \kappa_1 R$, while $m_1$ can extend up to $m_{1max}$. 
The final $\kappa_1$, $\kappa_2$ can be much larger than $\kappa$. 

Since the expression (\ref{integral-appendix}) is regularized with large but finite $R$, 
the summation and integration can be interchanged:
\be
\int rdr \, r'dr'\, J_m(\kappa r) J_m(\kappa r')
\sum_{m_1=-m_{1max}}^{+m_{1max}} J_{m_1}(\kappa_1 r) J_{m-m_1}(\kappa_2 r) J_{m_1}(\kappa_1 r') J_{m-m_1}(\kappa_2 r')\,.
\label{sum2}
\ee
Thanks to the properties of the Bessel functions, only $m_1$'s up to $min(\kappa_{1,2}r,\,\kappa_{1,2}r')$
are effectively contributing to this sum; for larger $m_1$ the Bessel functions strongly decrease. 
But $r,r' \le R$, which means that the limits on the summation can in fact be safely extended to the infinity.
Then, the sum of the product of four Bessel functions is treated in the following way:
\bea
&& \sum_{m_1=-\infty}^{+\infty} J_{m_1}(\kappa_1 r) J_{m-m_1}(\kappa_2 r) J_{m_1}(\kappa_1 r') J_{m-m_1}(\kappa_2 r') 
\label{sum4bessels}\\
&& = \sum_{m_1, m'_1=-\infty}^{+\infty} J_{m_1}(\kappa_1 r) J_{m-m_1}(\kappa_2 r) 
J_{m'_1}(\kappa_1 r') J_{m-m'_1}(\kappa_2 r')\cdot \delta_{m_1,m'_1}\nonumber\\
&& = \sum_{m_1, m'_1=-\infty}^{+\infty}  {1 \over 2\pi}\int_0^{2\pi} d\alpha e^{i(m_1-m'_1)\alpha} \,
J_{m_1}(\kappa_1 r) J_{m-m_1}(\kappa_2 r) J_{m'_1}(\kappa_1 r') J_{m-m'_1}(\kappa_2 r')\nonumber\\
&& = {1 \over 2\pi}\int_0^{2\pi} d\alpha 
\left[\sum_{m_1=-\infty}^{+\infty} e^{im_1\alpha} J_{m_1}(\kappa_1 r) J_{m-m_1}(\kappa_2 r)\right]
\left[\sum_{m'_1=-\infty}^{+\infty} e^{-im'_1\alpha} J_{m'_1}(\kappa_1 r') J_{m-m'_1}(\kappa_2 r')\right]\,.\nonumber
\eea
The first sum in the square brackets is calculates as follows:
\bea 
&&\sum_{m_1=-\infty}^{+\infty} e^{im_1\alpha} J_{m_1}(\kappa_1 r) J_{m-m_1}(\kappa_2 r) \nonumber\\
&&= {(-i)^m \over (2\pi)^2}\int d\phi_1\, d\phi_2\, e^{i\kappa_1 r \cos\phi_1 + i \kappa_2 r \cos\phi_2} 
\sum_{m_1=-\infty}^{+\infty}
e^{im_1\alpha + i m_1\phi_1 + i(m-m_1)\phi_2}\nonumber\\
&& =  {(-i)^m \over 2\pi}e^{im\alpha}\int d\phi_1\, e^{im\phi_1}e^{i\kappa_1 r \cos\phi_1 + i k_2 r \cos(\phi_1+\alpha)}\,.\label{sum_ajj}
\eea
The combination of angles and momenta inside the exponential can be expressed as 
\be
\kappa_1\cos\phi_1 + \kappa_2 \cos(\phi_1+\alpha) = |\bk_1+\bk_2|_{\alpha}\cos(\phi_1+\delta\phi)\,,
\ee
where
\be
\quad |\bk_1+\bk_2|_{\alpha} \equiv \sqrt{\kappa_1^2+\kappa_2^2 + 2\kappa_1 \kappa_2\cos\alpha}\,,\quad
\tan\delta\phi = {\kappa_2 \sin\alpha \over \kappa_1 + \kappa_2\cos\alpha}\,.
\ee
Geometrically, $|\bk_1+\bk_2|_{\alpha}$ is the norm of the sum of two vectors 
of moduli $\kappa_1$ and $\kappa_2$ and the relative azimuthal angle $\alpha$.
Therefore, (\ref{sum_ajj}) is
\be
\sum_{m_1=-\infty}^{+\infty} e^{im_1\alpha} J_{m_1}(\kappa_1 r) J_{m-m_1}(\kappa_2 r) = 
e^{im(\alpha-\delta\phi)}\, J_m(|\bk_1+\bk_2|_{\alpha} r)\,.
\ee
This expression can be viewed as a 2D generalization of the well-known addition formula for the Bessel functions
$$
\sum_{m_1=-\infty}^{+\infty} J_{m_1}(x)J_{m-m_1}(y) = J_{m}(x+y)\,.
$$
Now, the second sum differs only by $\alpha\to -\alpha$ (or, alternatively, complex conjugation)
and $r\to r'$. Therefore, the summation (\ref{sum4bessels}) is simplified to
\be
{1 \over 2\pi}\int_0^{2\pi} d\alpha\, J_m(|\bk_1+\bk_2|_{\alpha} r)\, J_m(|\bk_1+\bk_2|_{\alpha} r')\,.
\ee
Note that this integral involves only the Bessel functions of small order.
We now plug this expression in (\ref{sum2}) and get
\be
{1 \over 2\pi}\int_0^{2\pi} d\alpha\,
\left[\int rdr\, J_m(\kappa r)\, J_m(|\bk_1+\bk_2|_{\alpha} r)\right]
\left[\int r'dr'\, J_m(\kappa r')\, J_m(|\bk_1+\bk_2|_{\alpha} r')\right]
\ee
As usual, we extend the integration range in one of the integrals to infinity, which gives 
a delta-function, and then we use it on the second integral calculated up to $R$.
The original expression (\ref{integral-appendix}) then becomes
\be
{1 \over 2\pi} \int_0^{2\pi} d\alpha\,{\delta(\kappa-|\bk_1+\bk_2|_{\alpha}) \over \kappa} 
{R \over \pi \kappa} = {R \over 2\pi^2\kappa}\cdot {1\over \Delta}\,,\label{appendix-result}
\ee
where $\Delta$ is, as always, the area of the triangle with sides $\kappa$, $\kappa_1$, $\kappa_2$.

\end{document}